\documentstyle[prc,preprint,tighten,aps]{revtex}
\begin {document}
\draft
\title{
Pairing correlations in $N \sim Z$  $pf$-shell nuclei}
\author{K. Langanke$^{1,2}$, D.J. Dean$^3$, S.E. Koonin$^1$, and
P.B. Radha$^1$}
\address{
$^1$W.~K. Kellogg Radiation Laboratory, California 
Institute of Technology, Pasadena, California 91125\\
$^2$ Institute for Physics and Astronomy, Aarhus University, Denmark\\
$^3$Physics Division, Oak Ridge National Laboratory, P.O. Box 2008,
Oak Ridge, Tennessee 37831\\
}
\date{\today}

\maketitle

\begin{abstract}
\noindent
We perform
Shell Model Monte Carlo calculations to
study pair correlations in the ground states of $N=Z$ nuclei
with masses  $A=48-60$.
We find that $T=1$, $J^{\pi}=0^+$ proton-neutron correlations play an important,
and even dominant role, in the ground states of odd-odd $N=Z$ nuclei,
in agreement with experiment. By studying pairing in the ground states
of $^{52-58}$Fe, we observe that the isovector 
proton-neutron correlations decrease rapidly with increasing neutron excess.
In contrast,
both the proton, and trivially the neutron correlations
increase as neutrons are added. 
 
We also study the thermal properties and the temperature dependence of pair
correlations
for $^{50}$Mn and $^{52}$Fe as exemplars of odd-odd and even-even
$N=Z$ nuclei. While for $^{52}$Fe results are similar to those
obtained for other even-even nuclei in this mass range, 
the properties of $^{50}$Mn
at low temperatures are strongly influenced by isovector neutron-proton pairing.
In coexistence with these isovector pair correlations, 
our calculations also indicate an excess of isoscalar
proton-neutron pairing over the mean-field values.
The isovector neutron-proton correlations rapidly decrease
with temperatures and vanish for temperatures above
$T=700$ keV, while the isovector correlations among like nucleons
persist to higher temperatures.  Related to the 
quenching of the isovector proton-neutron correlations,
the average isospin
decreases from 1,
appropriate for the ground state, to 0
as the temperature increases.
\end{abstract}

\narrowtext

\section{Introduction}

Proton-rich radioactive ion-beam facilities offer the novel possibility
of exploring the structure
of nearly self-conjugate ($N \sim Z$) nuclei in the medium mass range 
$Z \lesssim 50$. Special interest will be devoted to understanding 
the proton-neutron
(pn) interaction, which 
has long been recognized to play
a particularly important role in $N=Z$ nuclei. The pn correlations can either
correspond to isovector ($T=1$) or isoscalar ($T=0$) pairs. Like proton-proton
(pp) and neutron-neutron (nn) pairing, isovector pn correlations in light to
medium mass nuclei are assumed to involve a proton-neutron pair in
time-reversed spatial orbitals, while the isoscalar correlations involve
mainly pairing between identical orbitals and spin-orbit partners \cite{Goodmann}.

The importance of pn correlations in
self-conjugate odd-odd nuclei is evident from the ground state spins
and isospins. In the $sd$-shell, odd-odd $N=Z$ nuclei
(with the exception of $^{34}$Cl)
have ground states with isospin $T=0$ and angular momenta $J > 0$,
pointing to the importance of isoscalar pn pairing between
identical orbitals. In contrast, self-conjugate $N=Z$ nuclei in the
medium mass range ($A > 40$)  have ground states with $T=1$
and $J^{\pi}=0^+$ (the only
known exception is $^{58}$Co) indicating the dominance of isovector
pn pairing.

Additional confirmation of the importance of $T=1$ pn-pairing 
in medium-mass odd-odd $N=Z$ nuclei is
an experiment that identified the $T=0$ and $T=1$ bands
in $^{74}$Rb \cite{Rudolph}. An isospin $T=1$, arising from isovector 
pn correlations, has been assigned
to the ground state rotational band in this nucleus.
At higher rotational frequency, or equivalently higher excitation energy,
a $T=0$ rotational band becomes energetically favored.

The competition of isovector and isoscalar
pn pairing has been extensively studied for $sd$-shell nuclei
\cite{Sandhu} and for nuclei at the beginning of the $pf$-shell \cite{Wolter}
using the Hartree-Fock-Bogoliubov (HFB) formalism. 
A major result 
\cite{Goodmann} is that $T=0$ pn correlations dominate $T=1$ 
correlations in the $N=Z$ nuclei studied;
$T=1$ pn pairing 
was never found to be important.
As mentioned above,
this is surprising since
the $T=1$ ground state isospin
of most odd-odd $N=Z$ nuclei with $A\ge40$ clearly points to
the importance of $T=1$ pn pairing in these nuclei.
It has been pointed out recently that HFB calculations 
for intermediate mass nuclei can exhibit nearly degenerate minima 
that may or may not involve
important $T=1$ pn correlations \cite{Faessler}.
However, these studies assume
that the $T=1$ pn pairing strength is larger than the nn and pp
pairing strengths.

Although HFB calculations have already pioneered the study of
pairing in $N=Z$ nuclei,
the method of choice to study pair correlations
is the interacting shell model.
Within the $sd$ shell \cite{Wildenthal} and at the beginning of the
$pf$-shell \cite{McGrory,Caurier}
the interacting shell model has proven to give an excellent
description of all nuclei, including the correct reproduction
of the spin-isospin assignments of self-conjugate $N=Z$ nuclei.
However, the conventional shell model using diagonalization techniques
is currently restricted to nuclei with masses $A \leq48$ due to computational
limitations. These limitations are overcome by
the recently 
developed Shell Model Monte Carlo (SMMC) 
approach \cite{Johnson,Lang}. Using this novel method,
it has been demonstrated \cite{Langanke1} that complete $pf$ shell 
calculations using the modified Kuo-Brown interaction well reproduce the
ground state properties of even-even and $N=Z$ nuclei with $A \leq 60$; 
for heavier nuclei an extension of the model space to 
include the $g_{9/2}$ orbitals is necessary.
Additionally the SMMC approach naturally allows the study of thermal
properties. First studies have been performed for several
even-even nuclei in the astrophysically interesting mass range $A=54-60$
\cite{Dean,Langanke2}.

In this paper we extend SMMC studies 
to detailed calculations of the 
$N=Z$ nuclei with $A=50-60$. The studies consider
all configurations within the complete $pf$-shell model space. Special 
attention is paid to isovector
and isoscalar pairing correlations in the ground states.
To elucidate the experimentally observed
competition between $T=1$ and $T=0$ correlations as a function of
excitation energy, we have also performed SMMC studies of the thermal
properties of an even-even and an odd-odd $N=Z$ nucleus
($^{52}$Fe and $^{50}$Mn, respectively).
In particular, we discuss the differences in
the thermal behavior of the pair correlations,
and other selected observables, in these nuclei.

\section{Model}

The SMMC approach was developed in Refs.
\cite{Johnson,Lang}, where the reader can find a detailed
description of the ideas underlying the method, its formulation,
and numerical realization. As the present calculations follow
the formalism developed and published previously, a very brief
description of the SMMC approach suffices here. 
A comprehensive review of the SMMC method and its applications can be found in
Ref. \cite{report}.

The SMMC method
describes the nucleus by a canonical ensemble at temperature  
$T=\beta^{-1}$ and employs a Hubbard-Stratonovich linearization 
\cite{Hubbard} of the  
imaginary-time many-body propagator, $e^{-\beta H}$, to express  
observables as path integrals of one-body propagators in fluctuating  
auxiliary fields \cite{Johnson,Lang}. Since Monte Carlo techniques  
avoid an explicit enumeration of the many-body states, they can be  
used in model spaces far larger than those accessible to conventional  
methods. The Monte Carlo results are in principle exact and are in  
practice subject only to controllable sampling and discretization  
errors. 
The notorious ``sign problem'' encountered in the Monte
Carlo shell model calculations with realistic interactions \cite{Alhassid}
can be circumvented
by a procedure suggested in Ref. \cite{Dean},  
which is based on an  extrapolation from a family of
Hamiltonians that is free of the sign problem
to the physical Hamiltonian. 
The numerical details of our calculation parallel those of Refs.
\cite{Langanke1,Dean}.

As we will show below, isovector pair correlations depend strongly
on the neutron excess in $N \sim Z$ nuclei,
so that a proper particle number
projection is indispensable for a meaningful study of these correlations.  
We stress that this important requirement is fulfilled
by the present SMMC approach, which uses a {\it canonical} expectation value
for all observables at a given temperature; i.e.,   
the proper proton and
neutron numbers of the nuclei are guaranteed by an appropriate number
projection \cite{Lang,report}.

The main focus of this paper is on pairing correlations 
in the three isovector $J^{\pi}=0^+$ channels and the isoscalar
proton-neutron correlations in the $J^{\pi}=1^+$ channel.
In complete $0 \hbar \omega$ shell model calculations.
the definition of the pairing strength is somewhat arbitrary. 
In this paper, we follow Ref. \cite{Langanke2}
in our description of pairing correlation 
and define
a pair of protons or neutrons
with angular momentum quantum numbers $(JM)$
by ($c=\pi$ for protons and $c=\nu$ for neutrons)
\begin{equation}
A_{JM}^\dagger (j_a,j_b) = 
\frac{1}{\sqrt{1+\delta_{j_a j_b}}}
\left[ c_{j_a}^\dagger c_{j_b}^\dagger \right]_{(JM)} ,
\end{equation}
where $\pi_j^\dagger$ ($\nu_j^\dagger$)
creates a proton (neutron) in an orbital with total spin $j$.
The 
isovector (plus sign)
and isoscalar (minus sign) proton-neutron pair operators are given by
\begin{equation}
A_{JM}^\dagger (j_a,j_b) = 
\frac{1}{\sqrt{2(1+\delta_{j_a j_b})}}
\left[ 
\nu_{j_a}^\dagger \pi_{j_b}^\dagger \pm
\pi_{j_a}^\dagger \nu_{j_b}^\dagger 
\right]_{(JM)}.
\end{equation}
With these definitions, we build up a pair matrix
\begin{equation}
M_{\alpha \alpha'}^J = \sum_M \langle A_{JM}^\dagger (j_a,j_b)
A_{JM} (j_c,j_d) \rangle  ,
\end{equation}
which corresponds to the calculation of the canonical ensemble average
of two-body operators like $\pi_1^\dagger \pi_2^\dagger \pi_3 \pi_4$.
The index $\alpha$ distinguishes the various possible $(j_a,j_b)$ 
combinations (with $j_a \geq j_b$).
The square matrix $M$ for the $pf$ shell has
dimension $N_J=4$ for $J=0$ and $N_J=7$ for $J=1$.
In Ref. \cite{Langanke2} 
the sum
of the eigenvalues of the matrix $M^J$ (its trace)
has been introduced as  a convenient overall measure for the strength
of pairs with spin $J$: 
\begin{equation}
P^J = \sum_{\beta} \lambda_{\beta}^J = \sum_{\alpha} M^J_{\alpha \alpha},
\end{equation}
where the $\lambda_\beta$ are eigenvalues of the matrix $M$.

An alternative often used to measure
the overall pair correlations
in nuclear wave functions, is in terms 
of the BCS pair operator
\begin{equation}
\Delta_{JM}^\dagger = \sum_{\alpha}  A_{JM}^\dagger
(\alpha).
\end{equation}
The quantity $\sum_M \langle \Delta_{JM}^\dagger \Delta_{JM} \rangle$
is then a
measure of the number of nucleon pairs with spin $J$.
We note that, for the results discussed in this paper,
the BCS-like definition for the overall pairing
strength yields the same qualitative results for the pairing content
as the definition (4).
Some SMMC results for BCS pairing in nuclei in the mass range $A=48-60$
are published in Refs. \cite{Langanke1,Langanke2,report,Engel}.

With our definition (4) the pairing strength is non-negative, and indeed
positive, at the 
mean-field level. The 
mean-field pairing
strength, $P_{\rm mf}^J$, can be defined as in (3,4),
but replacing the expectation values of the two-body matrix
elements in the definition of $M^J$ by
\begin{equation}
\langle c_1^\dagger c_2^\dagger c_3 c_4 \rangle \rightarrow
n_1 n_2 \left( \delta_{13} \delta_{24} - \delta_{23} \delta_{14} \right) ,
\end{equation}
where $n_k = \langle c_k^\dagger c_k \rangle$ is the occupation number
of the orbital $k$. This mean-field value provides a baseline against
which true pair correlations can be judged.

\section{Results}

Our SMMC studies for self-conjugate nuclei with $A=48-60$
have been performed using the modified
Kuo-Brown KB3 residual interaction \cite{KB3}. Some results of these
studies for observables like ground state energies and total
Gamow-Teller, $B(M1)$, and $B(E2)$  strengths have already been presented
in Ref. \cite{Langanke1}. As for other $pf$-shell nuclei, the SMMC results
for the self-conjugate nuclei
are generally in very good agreement with data.

As is customary in shell model studies, the Coulomb interaction
has been neglected, which we believe is a justified approximation
in this mass range. Thus, our shell model Hamiltonian is isospin-invariant
and, as a consequence, there are symmetries in the pairing strengths
of the three isovector $J^{\pi}=0^+$ channels. For even-even $N=Z$ nuclei,
$P^{J=0}$ is 
identical for pp, pn and nn
pairing.  In odd-odd self-conjugate nuclei (with ground state isospin
$T=1$), 
the equality of
the pp and nn channels remains,
but the pn part of the isovector multiplet
$P^{J=0}$ can differ from the other two 
components 
($P^{J=0}_{pp}=P^{J=0}_{nn} \ne P^{J=0}_{pn}$). 
At the mean-field level, the three components of the isovector
pairing multiplet are identical for both odd-odd and even-even $N=Z$ nuclei.

\subsection{Pairing in the ground states of self-conjugate $pf$-shell nuclei}

Our SMMC calculations for the even-even nuclei
have been performed at finite temperatures
$T=0.5$ MeV, 
which has been found sufficient in previous studies 
to guarantee cooling into the ground state.
The odd-odd nuclei were studied at $T=0.4$ MeV. Since the latter have
experimentally a low-lying excited state
with an excitation energy of about 0.2 MeV, our ``ground state''
calculations for these nuclei corresponds to a mixture of the ground 
and first excited states.

As expected, we calculate vanishing isospin and
angular momentum expectation values ($\langle T^2 \rangle$ and 
$\langle J^2 \rangle$, respectively) for the ground states of the even-even
nuclei. For the odd-odd nuclei our calculations yield isospin
expectation values $\langle T^2 \rangle = 2.2\pm0.3$ for $^{50}$Mn, 
and $1.8\pm0.2$ for $^{54}$Cu,
in good agreement with experiment, as both $^{50}$Mn
and $^{54}$Co have a $T=1$ ground state, so that $T(T+1)=2$.
For $^{58}$Cu we find
$\langle T^2 \rangle = 1.4\pm0.2$, while the experimental level spectrum 
has a $T=0$ $1^+$
ground state and a $T=1$ $0^+$ first excited state at $E_x=0.2$ MeV.
The error bars in our calculations for the angular momentum expectation
values 
($\langle J^2 \rangle = -1.4\pm4.5$ for $^{50}$Mn, $-7.0\pm7.5$
for $^{54}$Co and $3.0\pm4.0$ for $^{58}$Cu) prohibit meaningful
comparison with experiment. We note that the SMMC also
reproduces the $T=1$ isospin of the $^{62}$Ga and 
$^{74}$Rb ground states
(using the $p,f_{5/2},g_{9/2}$ model space \cite{Dean96}).
Detailed
calculations of the pairing  in these two nuclei will be presented in 
\cite{Dean96}.

We have calculated the isovector
and isoscalar  pairing strengths in the ground states of the self-conjugate
nuclei with $A=48-60$ using the definition (4). 
The results are presented in Fig. 1,
where they are also compared to the mean-field values derived
using Eq. (6).  
Discussing the isovector $J=0$ pairing channels first,
Fig. 1 shows an excess of pairing correlations
over the mean-field values.
For the even-even nuclei, this excess represents the well-known
pairing coherence in the ground state. 
In addition, Fig. 1 exhibits
a remarkable staggering in the $J=0$ pp 
and nn pairing channels
($P^{J=0}_{pp} = P^{J=0}_{nn}$) and in the pn
pairing channel ($P^{J=0}_{pn}$) when comparing neighboring even-even
and odd-odd self-conjugate nuclei. In the latter, the isovector pn pairing
clearly dominates the pp and nn pairing
and is always significantly 
larger than in the neighboring
even-even $N=Z$ nuclei. In contrast, the 
like-nucleon pairing 
is noticeably reduced in the odd-odd nuclei
relative to the values in the neighboring even-even nuclei.

The odd-even staggering is not visible in the total $J=0$
pairing strength, 
\begin{equation}
P^{J=0}_{tot}= 
P^{J=0}_{pp}+P^{J=0}_{nn} + P^{J=0}_{pn},
\end{equation} 
as can be seen in Fig. 1.
Although the excess of
$P^{J=0}_{tot}$ over the mean-field value is significant, it is 
about equal for the ``open shell'' nuclei
$^{48}$Cr, $^{50}$Mn, $^{52}$Fe and $^{60}$Zn. Towards
the $N=28$ shell closure the excess decreases and becomes a minimum
for the double-magic nucleus $^{56}$Ni. In fact, the  
excess of
$P^{J=0}_{tot}$ over the mean-field value is only $0.42\pm0.1$ 
in $^{56}$Ni (or about $13\%$
of $P^{J=0}_{tot}$), while it is $2.1\pm0.1$ in $^{48}$Cr (or about $350\%$).
We thus conclude that the change in the total pairing strength
in the $N=Z$ nuclei is governed by shell effects, but that there is a
significant redistribution of strength between the like-
and unlike-pairs in going from even-even to odd-odd nuclei,
with pn pairing favored in the latter.
 
Our calculations indicate that the $J=1$ pn channel is the most
important isoscalar pairing channel in the ground states. 
As is shown in Fig. 1, there is a modest excess of $J=1$ isoscalar pn pairing
over the mean-field values in all nuclei studied. The calculations
indicate a slight even-odd staggering in the pairing excess, with the excess
being larger in the even-even nuclei. Apparently
the strong isovector pn pairing decreases not only
the isovector pairing between like nucleons, but also the isoscalar
pn pairing. As in the isovector channels, the excess of isoscalar
pairing is strongly decreased close to the $N=28$ shell closure,
where the nuclei become spherical. It is well-known that isoscalar $J=1$
pn pairing is important in deformed nuclei like $^{48}$Cr.
We note that 
within the uncertainties of the calculation, 
our studies 
do not show any pairing excess above the mean-field values 
in the $J=3$, $5$, and
$7$ isoscalar channels.  

It is interesting to compare
the present SMMC results 
for the isovector pairing strength 
with those of
a simple seniority-like model with an isospin-invariant
pairing Hamiltonian \cite{Engel}. One finds
that the magnitude of the isovector
pairing correlations is smaller in the SMMC studies than in the
simple pairing model, as in the realistic shell model these
correlations compete with other nucleonic correlations (e.g.
the isoscalar pairing, which had not been considered in Ref. \cite{Engel}).
However, it is remarkable that the simple pairing model reproduces
the odd-even staggering seen in the
SMMC studies. 

Using the HFB approach, Wolter {\it et al.} have studied 
pairing in $^{48}$Cr, restricting themselves to considering
isovector and isoscalar pairing separately \cite{Wolter}. 
These authors find
the isoscalar pairing mode to be considerably stronger than the
isovector \cite{Wolter}. This finding is not supported
by our SMMC calculation; it might be caused by the fact that the HFB
solutions had not been projected on the appropriate ground state
angular momentum. In fact, the HFB solutions have 
$\langle J^2 \rangle \gg 0 $,
making the presence of aligned pairs necessary. The SMMC calculations,
which have 
$\langle J^2 \rangle \approx 0 $,
do not show the importance
of aligned pairing in the $^{48}$Cr ground state. 

To investigate
how the various pairing strengths change if the proton-neutron
symmetry of the $N=Z$ nuclei is broken
by adding additional neutrons,
we have performed a series of SMMC
studies for the iron isotopes $^{52-58}$Fe; the results are shown
in Fig. 2. The striking result is that the excess
of isovector pn pairing over the mean-field values is decreased 
drastically upon adding neutrons and has practically 
vanished in $^{56,58}$Fe. 
In contrast
the excess in both $J=0$ pp and nn pairing is increased
by moving away from charge symmetry. 
At the mean-field level pp pairing
is virtually unchanged through an isotope chain,
while adding neutrons increases 
the nn pairing.   
The excess in total pairing strength $P^{J=0}_{tot}$
within the isotope chain increases only slightly as neutrons are added.
Fig. 2 also shows that the excess of $P^{J=1}_{pn}$
pairing
decreases with neutron excess. However, this decline is less dramatic
than for $P^{J=0}_{pn}$  and it appears that in nuclei
with neutron excess, isoscalar $(J=1$) pn correlations are
more important than isovector. This finding
is in agreement with the observation that isoscalar pn
pairing is mainly responsible for the quenching of the Gamow-Teller 
strength \cite{Gamow},
as those SMMC investigations have been performed for nuclei with 
$N > Z$.

Note that $^{54}$Fe is exceptional as
it is magic in the neutron number ($N=28$).
As a consequence, the excess of nn pairing,
and also of isovector and isoscalar pn pairing, is low
compared to the other isotopes.

Our SMMC results for pairing correlations as function
of neutron excess thus yield the following simple picture.
Adding neutron pairs apparently increases the collectivity of the
neutron condensate so that there are fewer neutrons available
to pair with protons. As a result, protons pair more often with other
protons, in this way increasing the proton collectivity, although the
total number of protons, of course, remains unchanged. 

Based on the results of their simple pairing model, the authors of Ref.
\cite{Engel} come to the same conclusions. In fact,
the SMMC results for the changes of isovector pairing
within an isotope chain again agree well with the simple
pairing model. However, the ground state
in the latter is nearly
a product of pp and nn condensates (in the limit of large neutron excess), 
while
protons and neutrons 
in the realistic SMMC calculations 
still couple via
isoscalar (mainly $J=1$) correlations.

\subsection{Thermal properties of $^{50}$Mn and $^{52}$Fe}
 
To study the thermal properties of odd-odd and even-even $N=Z$ nuclei
we have performed SMMC studies of $^{50}$Mn and $^{52}$Fe at selected
temperatures $T \leq 2$ MeV; 
the results are presented in Figs. 3 and 4.
As we will show in the following, the thermal
properties of the two nuclei are dominated 
by the three isovector $J=0$ and the isoscalar $J=1$ proton-neutron
correlations; differences between the two nuclei can be traced
to differences in the thermal behavior of these correlations.

We note again that the three isovector $J=0$ pairing correlations are identical
in even-even $N=Z$ nuclei.
As discussed above, these correlations show a strong coherence and
dominate the ground state properties of $^{52}$Fe.
The temperature dependence of the 
$J=0$ pp correlations in $^{52}$Fe is very similar to those
of the other even-even iron isotopes $^{54-58}$Fe, which have been discussed
in Refs. \cite{Dean,Langanke2}. 
As in the other iron isotopes, the SMMC
calculations predict a phase transition 
in a rather narrow temperature interval around $T=1$ MeV,
where the  $J=0$ pairs break. Due to $N=Z$ symmetry, this phase
transition also occurs in the $J=0$ pn channel.
This behavior is different from the iron isotopes with neutron excess.
There the pn $J=0$ correlations have only 
a small excess at low temperatures,
where $J=0$ pairing among like nucleons dominates, and this excess actually
increases slightly when the like-pairs break \cite{Langanke2}.   
We also observe that 
the excess in the
isoscalar $J=1$ pn correlations is about constant at temperatures
below 2 MeV. Thus, these correlations persist to higher temperatures
than the isovector $J=0$ pairing, as has already been pointed out in
Ref. \cite{Langanke2}.

The pairing phase transition is accompanied by a rapid increase in
the moment of inertia,
$I=\langle J^2 \rangle /(3T)$ \cite{Langanke2}, 
and a partial unquenching (orbital part) of the total M1 strength.
The total Gamow-Teller strength also unquenches partially at the
phase transition, related to the fact that the vanishing of the $J=0$
pn
correlations reduces the quenching for Gamow-Teller transitions between
identical orbitals (mainly the $f_{7/2}$ 
proton to $f_{7/2}$ neutron transitions).
The residual quenching of the Gamow-Teller strength at temperatures above
the phase transition (the single particle value is 13.1 calculated from
the various ground state occupation numbers)
is caused by the isoscalar pn correlations,
which persist to higher temperatures (see Fig. 3). We note that the temperature
evolution of the Gamow-Teller transition is different in $^{52}$Fe than in
the iron isotopes with neutron excess \cite{Dean,Langanke2}. 
In the latter, $J=0$ pn
correlations do not show any significant excess over the mean-field values, and
in particular, do not exhibit the phase transition 
as in $^{52}$Fe. As a consequence,
the Gamow-Teller strength in nuclei with neutron excess is roughly constant
across the pairing phase transition, without any noticeable
unquenching.

As required by general thermodynamic principles,
the internal energy
increases monotonically with temperature. 
The heat capacity $C(T)=dE/dT$ 
is usually associated with a level density parameter $a$
by $C(T)=2a(T) T$.
As is typical for even-even 
nuclei \cite{Dean} $a(T)$
increases from $a=0$ at $T=0$
to a roughly constant value at temperatures above the phase transition.
We find $a(T) \approx 5.3\pm1.2$ MeV$^{-1}$ at $T \geq 1$ MeV, in 
agreement with the empirical value of 6.5 MeV$^{-1}$ \cite{Thielemann}.
At higher temperatures, $a(T)$ must decrease due to
the finite model space of our calculation. 
The present temperature grid is not fine enough to determine whether
$a(T)$ exhibits a maximum related to the phase transition, as suggested
in \cite{Dean}.

As expected for an even-even nucleus, $\langle T^2 \rangle$
is zero at low temperatures and then slowly increases with temperature
as higher isospin configurations are mixed in.

The thermal properties of $^{50}$Mn (Fig. 4) show
some distinct differences from $^{52}$Fe, which we believe are typical for
odd-odd $N=Z$ nuclei in this mass range. As already stressed in the last
section, $J=0$ pn correlations dominate 
the ground state properties. With increasing temperature
these correlations  decrease rapidly and steadily
and have already dropped to the mean-field
value at $T=0.8$ MeV. (The fact that the correlations actually become
slightly negative is unphysical and due
to uncertainties in the extrapolation
required to avoid the sign problem \cite{Alhassid}.
We have verified the qualitative results of our calculation for
an isospin invariant pairing+quadrupole Hamiltonian which does not exhibit
the sign problem \cite{Zheng}.)
The $J=0$ pp (and the identical nn) correlations
show the same phase transition near $T=1$ MeV, as in $^{52}$Fe
and the other even-even nuclei \cite{Dean,Langanke2}. In contrast
however,
the excess of  the $J=0$ correlations 
between like nucleons 
in $^{50}$Mn 
is noticeably smaller at low temperatures. 
As in $^{52}$Fe, the moment of inertia of $^{50}$Mn
increases drastically
when the $J=0$ pairs break.

We observe that the isoscalar $J=1$ pairing in $^{50}$Mn is rather
similar to that in $^{52}$Fe. In particular the excess of these
correlations is roughly constant 
in the temperature range where the
isovector $J=0$ correlations vanish and persists to higher temperatures than
the excess in the isovector correlations.

Related to the rapid decrease of isovector pn pairing
with temperature, the isospin expectation value drops from 
$\langle T^2 \rangle =2$ at $T \le 0.5$ MeV to
$\langle T^2 \rangle \approx 0.2\pm 0.2$ at $T = 1.0$ MeV;
the unpaired proton and neutron apparently recouple from an isovector
$J=0$ coupling to an isoscalar $J=1$ coupling.
At temperatures above $T=1$ MeV (i.e., above the phase transition for
like-nucleon pairs), the temperature dependences of the isospin expectation
values in $^{50}$Mn and $^{52}$Fe are similar.

The temperature dependence of the energy $E=\langle H \rangle$ in $^{50}$Mn
is significantly different than that in the even-even nuclei.
As can be seen in Fig.4, $E$ increases approximately linearly
with temperature, corresponding to a constant heat capacity
$C(T) \approx 5.4\pm1$ MeV$^{-1}$; 
the level density parameter decreases like $a(T) \sim T^{-1}$
in the temperature interval between 0.4 MeV and 1.5 MeV.
We note that the
same linear increase of the energy with temperature is observed in SMMC
studies of odd-odd $N=Z$ nuclei performed with a pairing+quadrupole
hamiltonian \cite{Zheng} and thus appears to be generic for self-conjugate
odd-odd $N=Z$ nuclei.

>From the discussion above, we conclude that
the main effect of heating the nuclei $^{50}$Mn
and $^{52}$Fe to temperatures around 1.5 MeV is to release the pairing
energy stored in the isovector $J=0$ correlations. From Fig. 1 
(i.e., $P^J_{\rm tot}$) we expect
that this pairing energy is about the same for both nuclei. This is in
fact confirmed by our calculation:
Using a linear extrapolation to zero
temperature, we find an internal excitation energy
of about 8.6 MeV at $T=1.6$ MeV in $^{50}$Mn, which agrees nicely with
the value for $^{52}$Fe, if we approximate $E(T=0)=E(T=0.5$ MeV). 
The apparent difference between the two nuclei is that, with increasing
temperature, a strong pairing gap in the three isovector $J=0$ channels
has to be overcome in $^{52}$Fe, while no such strong gap 
in the dominant isovector pn channel appears to exist in the
odd-odd $N=Z$ nucleus $^{50}$Mn. 

Associated with the decrease of the isovector pn
correlations, the Gamow-Teller strength in $^{50}$Mn
unquenches with heating
to $T=1$ MeV. This is noticeably different than in even-even nuclei,
where the Gamow-Teller strength is roughly constant
at temperatures below the pairing phase transition. 
We stress that the $B(GT)$ strength, however,
remains noticeably quenched even above $T=1$ MeV
due to the persistence of the isoscalar pn correlations;
the mean-field value for the Gamow-Teller strength increase only slightly
from 11.1 at $T=0.4$ MeV to 11.6
at $T=2$ MeV. 

The $B(M1)$ strength in $^{50}$Mn
decreases when the isovector pn correlations
vanish. It unquenches at
$T\ge0.8$ MeV, related to the breaking
of the pp and nn pairs, as shown in \cite{Langanke2}. 
As in even-even nuclei, the $B(M1)$ strength in $^{50}$Mn remains
quenched even for $T\approx 2$ MeV (the mean-field value
is 33.6 $\mu_N^2$ at $T=2$ MeV)
due to the
persisting isoscalar pn pairing.

\section{Conclusion}

We have studied pairing correlations in self-conjugate
nuclei in the middle of the $pf$ shell;
our calculations are the first to take all relevant two-nucleon
correlations into account. 
Our study is based on SMMC
calculations of these nuclei within the complete $pf$ shell employing the
realistic Kuo-Brown interaction KB3. Several results of our investigation
are noteworthy.

The isovector $J=0$ pairing correlations show a significant staggering
between odd-odd and even-even $N=Z$ nuclei. While the three isovector
channels have identical strengths in even-even $N=Z$ nuclei, 
the total isovector pairing strength is strongly redistributed
in odd-odd self-conjugate nuclei, with a strong enhancement of the 
proton-neutron correlations. This redistribution manifests itself in
a significantly different temperature dependence of observables like
the $GT$ and $B(M1)$ strengths and the internal energy.

The importance of isovector proton-neutron correlations decrease drastically
if neutrons are added. These additional neutrons increase the coherence
among the neutron condensate, thus making less neutrons available
for isovector proton-neutron correlations.
At the same time, the correlations among protons also increase
if neutrons are added. Our calculations indicate that in nuclei
with large neutron excess, isoscalar ($J=1$) proton-neutron correlations
dominate over isovector proton-neutron pairing.

We have studied the temperature dependence of the pairing correlations
and of selected observables for $^{50}$Mn and $^{52}$Fe.
While the even-even $N=Z$ nucleus $^{52}$Fe shows the same qualitative trends
as other even-even nuclei in this mass region (including a pairing
phase transition at temperatures near $T=1$ MeV), the results for
the odd-odd nucleus $^{50}$Mn differ in some interesting aspects. 
While the proton-proton and neutron-neutron correlations (although
much weaker than in even-even nuclei)
show a phase transition near $T=1$ MeV, the dominant 
$J=0$ proton-neutron correlations decrease steadily with increasing temperature.
As a consequence the internal energy increases linearly with temperature,
indicating that there is no pairing gap 
in the $J=0$ proton-neutron channel
to be overcome.

\acknowledgements

This work was supported in part by the National Science Foundation,  
Grants No. PHY94-12818 and PHY94-20470. Oak Ridge National Laboratory
is managed by Lockheed Martin Energy Research Corp. for the U.S. Department
of Energy under contract number DE-AC05-96OR22464.
We are grateful to J.~Engel, B.~Mottelson and  P.~Vogel 
for helpful discussions. DJD acknowledges an E.P. Wigner Fellowship
from ORNL.
Computational
cycles were provided by
the Concurrent Supercomputing Consortium at Caltech
and by the VPP500,
a vector parallel processor at  
the RIKEN supercomputing facility; we thank Drs. I. Tanihata and
S. Ohta for their assistance with the latter.

\narrowtext
\begin{figure}
\caption{Pairing correlations $P^J$ in the ground state of the $N=Z$ nuclei
with masses $A=48-60$. The full circles show the SMMC results, while
the open circles are the mean-field values. 
$P^{J=0}_{\rm tot}$ gives
the sum of the three isovector pair correlations in the $J=0$ channel.
Note that $P^{J=0}_{pp} = P^{J=0}_{nn}$ for $N=Z$ nuclei.}
\label{fig1}
\end{figure}

\narrowtext
\begin{figure}
\caption{Pairing correlations $P^J$ 
in the ground states of the even iron isotopes
$^{52-58}$Fe.
The full circles show the SMMC results, while
the open circles are the mean-field values. 
$P^{J=0}_{\rm tot}$ gives
the sum of the three isovector pair correlations in the $J=0$ channel.}
\label{fig2}
\end{figure}

\narrowtext
\begin{figure}
\caption{Thermal properties of $^{52}$Fe. The SMMC results are shown
with error bars, while the lines indicate the mean-field values
for the respective pair correlations.}
\label{fig3}
\end{figure}

\narrowtext
\begin{figure}
\caption{Thermal properties of $^{50}$Mn. The SMMC results are shown
with error bars, while the lines indicate the mean-field values
for the respective pair correlations.}
\label{fig4}
\end{figure}

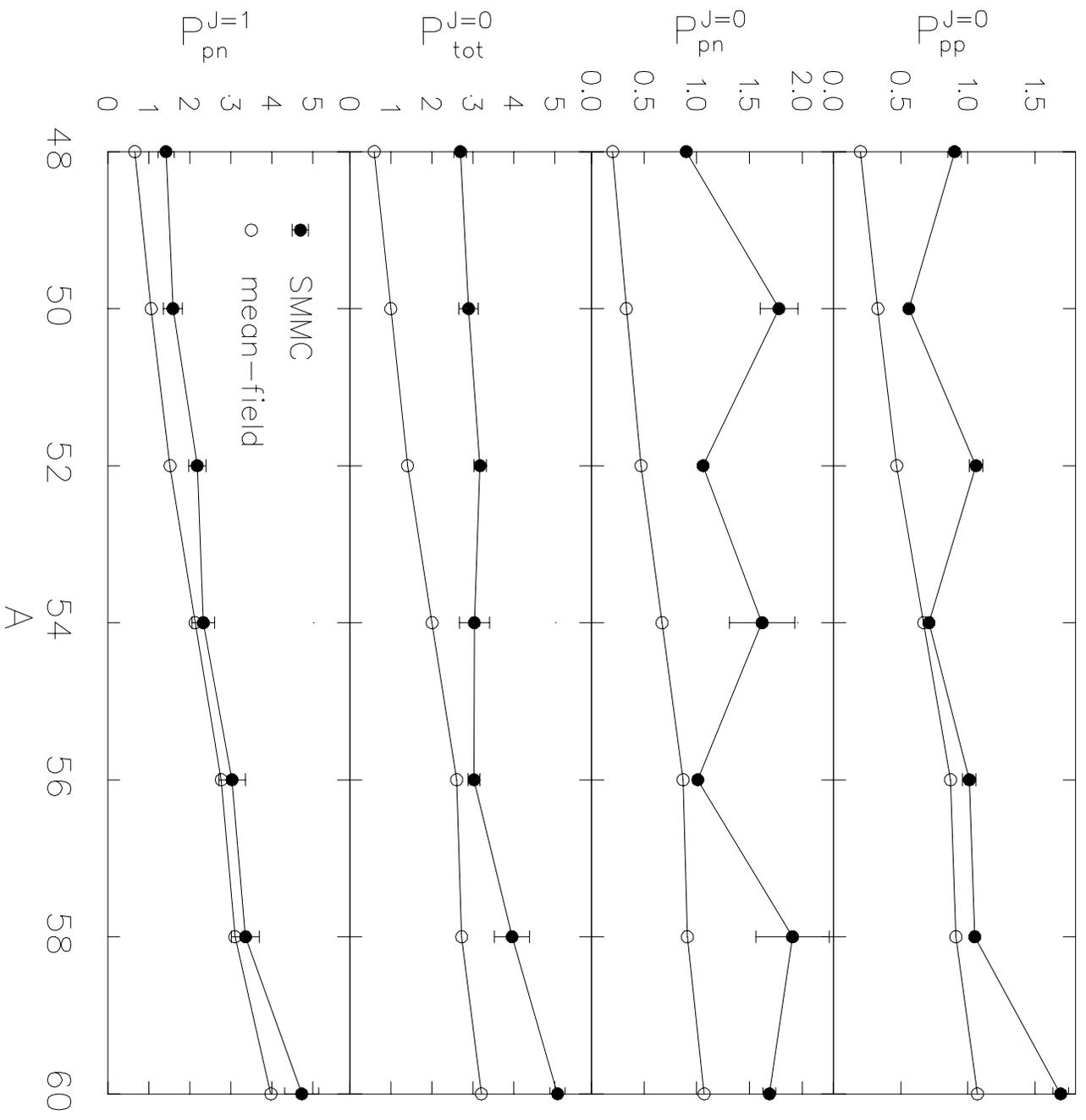
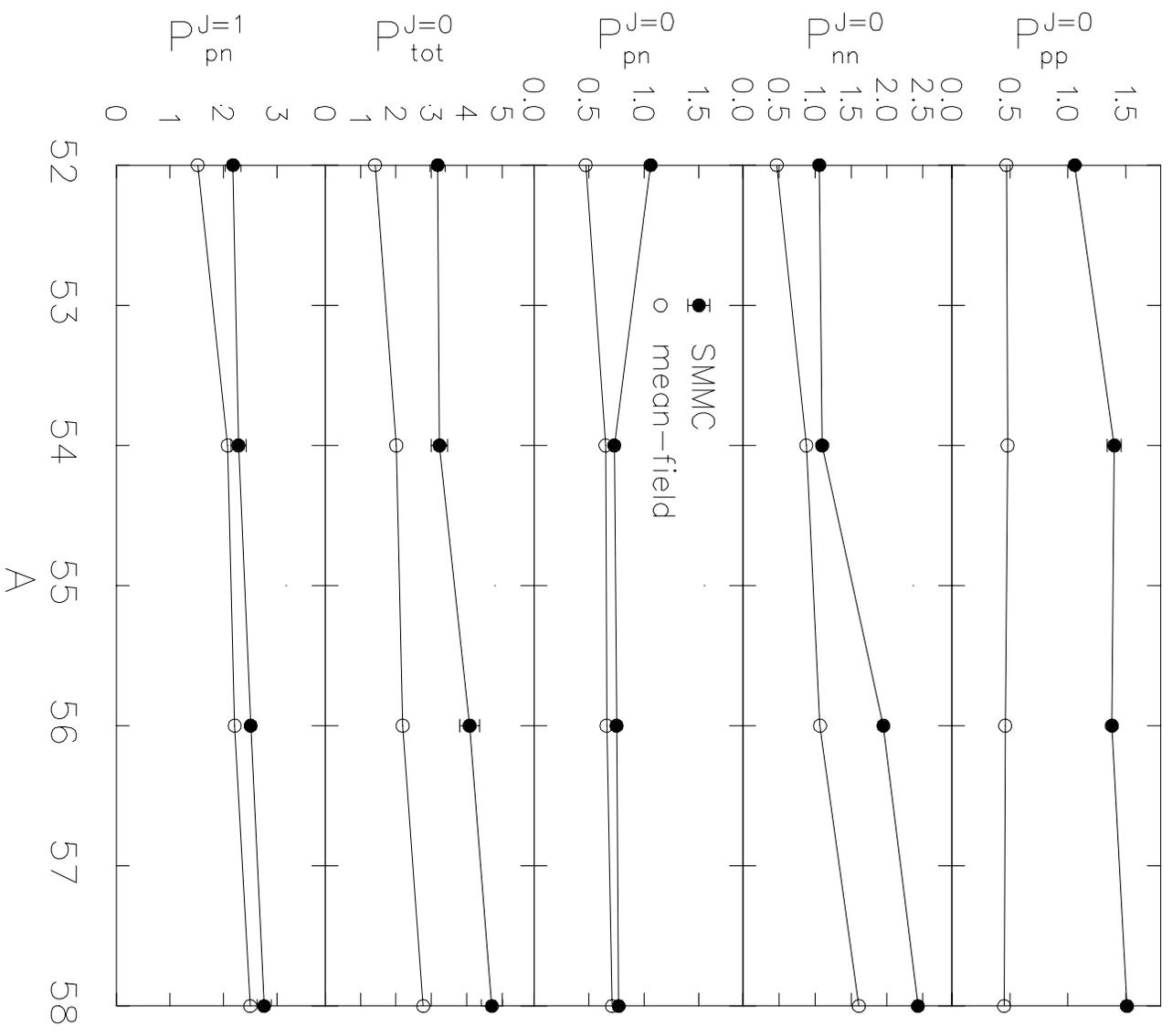
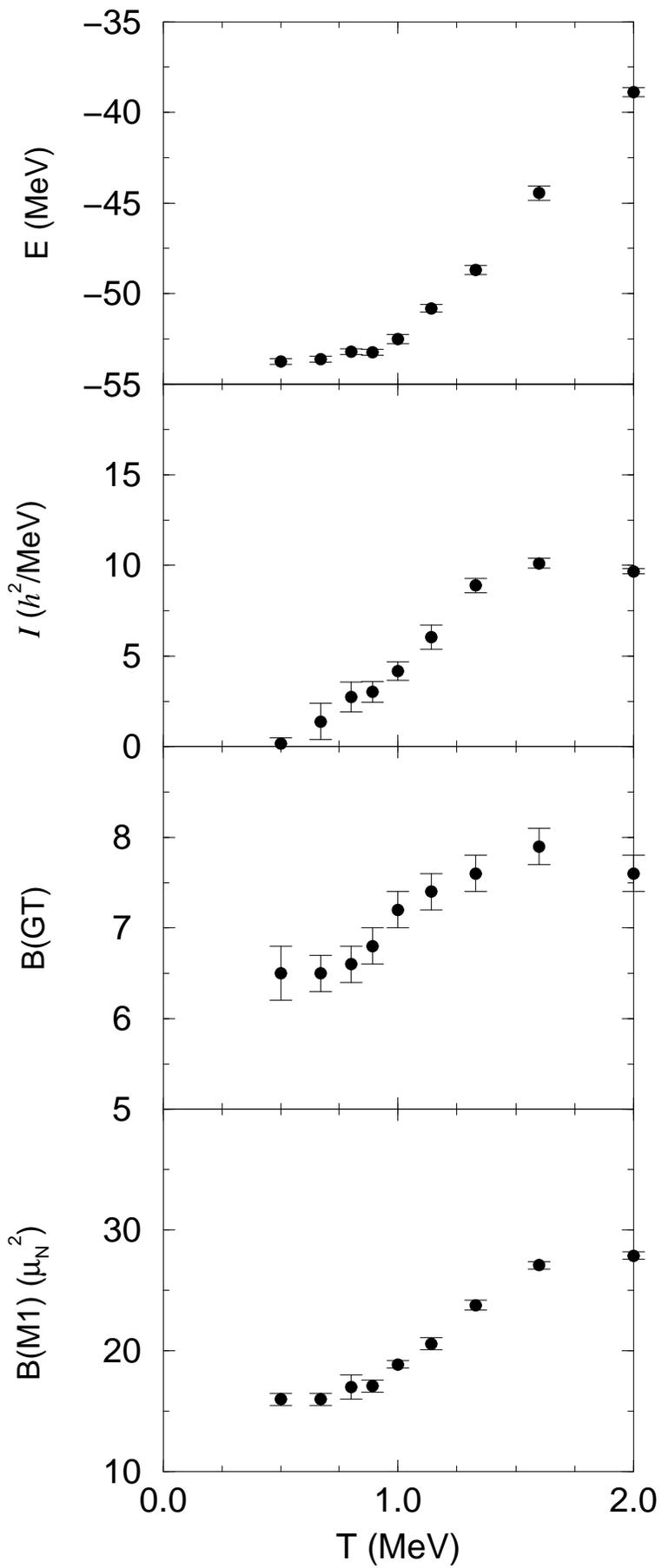
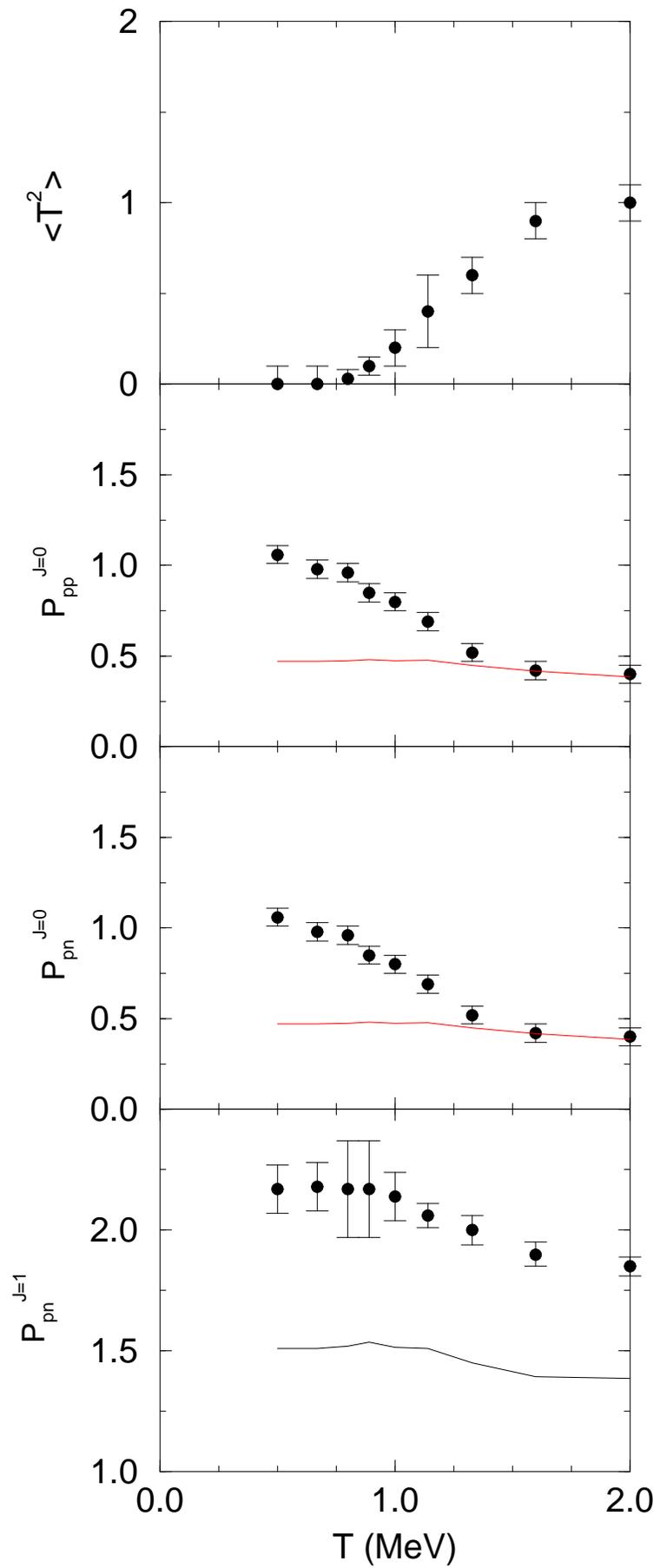
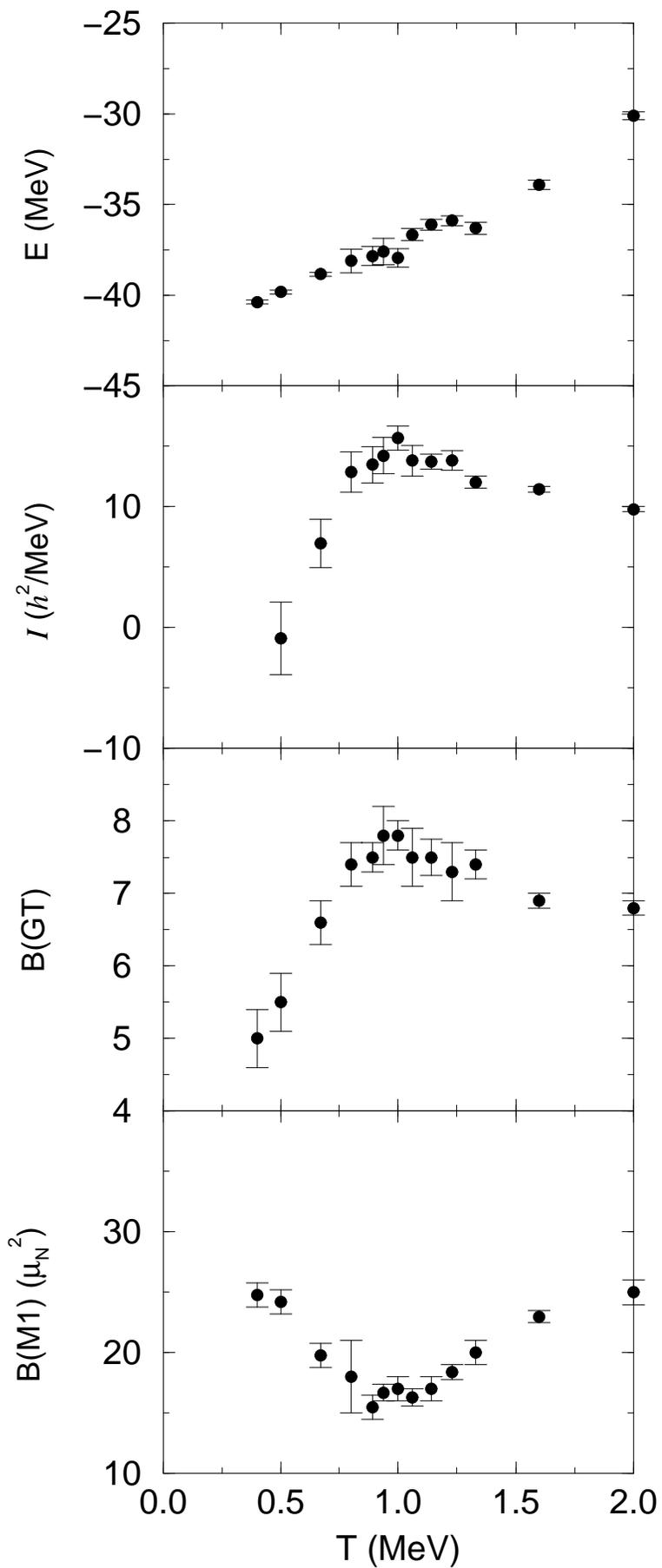
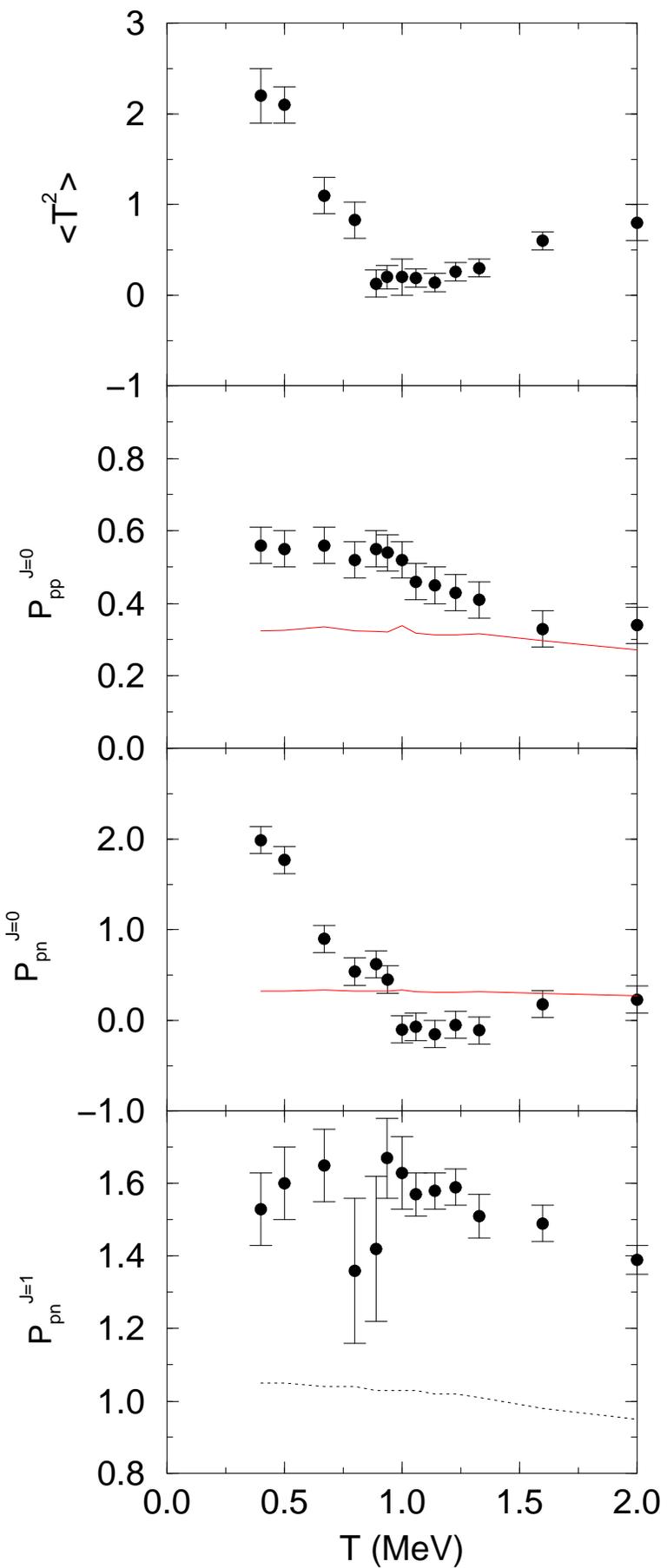
\end{document}